\documentclass{aa}  
\usepackage{graphicx}
\graphicspath{ {./images/} }
\usepackage{txfonts}
\usepackage{amsmath}
\usepackage[colorlinks = true,
            linkcolor = blue,
            urlcolor  = blue,
            citecolor = blue,
            anchorcolor = blue]{hyperref}
\begin{document} 

   \title{Using SRG/eROSITA to estimate soft \\  
     proton fluxes at the ATHENA detectors}


\author{E.~Perinati
\inst{1}
\and
M.~Freyberg
\inst{2}
\and
M.~C.~H.~Yeung
\inst{2}
\and
C.~Pommranz
\inst{1}
\and
B.~Heß
\inst{1}
\and
S.~Diebold
\inst{1}
\and
C.~Tenzer
\inst{1}
\and
A.~Santangelo
\inst{1}
}

{
\institute{Institut für Astronomie und Astrophysik, Eberhard Karls Universität Tübingen, Sand 1, 72076 Tübingen, Germany\\
\email{emanuele.perinati@uni-tuebingen.de}
\and
Max-Planck-Institut für extraterrestrische
Physik, Gießenbachstraße 1, 85748 Garching, Germany}

    \date{Received ? , accepted ?}

 
  \abstract
    {Environmental soft protons have affected the performance of the X-ray detectors on board Chandra and XMM-Newton, and they pose a threat for future
      high energy astrophysics missions with larger aperture, such as ATHENA.}
   {We aim at estimating soft proton fluxes at the ATHENA detectors
     independently of any modelisation of the external fluxes in the space environment.}
   {We analysed the background data measured by eROSITA on board SRG, and with the help of simulations we defined a range of values for the potential count-rate of quiet-time soft protons focused through the mirror shells. We used it to derive an estimate of soft proton fluxes at the ATHENA detectors, assuming ATHENA in the same L2-orbit as SRG.} 
   {The scaling, based on the computed proton transmission yields of the optics and optical/thermal filters of eROSITA and ATHENA, indicates that the soft proton induced WFI and X-IFU backgrounds could be expected close to the requirement.}
    {No soft proton fluxes detrimental to the observations have been suffered by eROSITA during its all-sky survey in orbit around L2. Regardless of where ATHENA will be placed (L1 or L2), our analysis suggests that increasing somewhat the thickness of the WFI optical blocking filter, e.g.\ by $\sim$30\%, would help to reduce the soft proton flux onto the detector, in case the planned magnetic diverters perform worse than expected due to soft proton neutralisation at the mirror level.     
    }
     

   \maketitle
%

   \section{Introduction}

The in-orbit experience of Chandra and XMM-Newton revealed that environmental soft protons   
represent a hazard for X-ray telescopes in space, as X-ray optics are able to focus them onto the detectors. Soft protons can contribute to degrade the detector performance inducing both radiation damage and instrumental background. The possible contamination by soft protons is especially a concern for the future \emph{Advanced Telescope for High ENergy Astrophysics} (ATHENA) \citep{Nandra2013} planned by ESA, mainly because of the large aperture of its Silicon Pore Optics (SPO) \citep{Collon2023}. Our laboratory tests on an SPO sample confirmed its ability to forward scatter soft protons in a broad range of energies from a few tens of keV to a few hundreds of keV \citep{Amato2021}. However, for ATHENA radiation damage from focused soft protons is not a major issue, the Wide Field Imager (WFI) \citep{Meidinger2020} being a back-illuminated DEPFET-based device integrated onto a 450 $\mu$m thick fully depleted silicon bulk where soft protons would be stopped within a few micrometers underneath the passivated surface; and the X-ray Integral Field Unit (X-IFU) \citep{Barret2018} being a thermal detector where the signal carriers are phonons rather than electron-hole pairs. On the contrary, the background of both instruments could be severely affected by focused soft protons. Previous attempts to assess soft proton fluxes at the ATHENA focal plane have depicted an adverse scenario where, even during quiet-time, i.e.\ in absence of solar particle events (SPE), a contribution to the instrumental background largely above the requirement of 5·10${^{-4}}$ cts/cm${^{2}}$/s/keV in the 2-7 keV band could be expected for an interplanetary orbit around the second Lagrangian point L2 \citep{Fioretti2018}. This has led to the adoption of ad-hoc magnetic diverters as a countermeasure to prevent soft protons from reaching the focal plane and to even consider the possibility to change the ATHENA orbit from L2 to L1. Indeed a similar investigation is currently ongoing for an orbit around L1, where the radiation environment would possibly be more benign than L2, yet it is anticipated that also there the magnetic diverters would be essential to meet the requirement.

Unfortunately, more recently our laboratory tests applying a magnetic field provided evidence that soft protons with an energy below 100 keV scattered off an SPO sample at grazing angles undergo neutralisation. This effect has been known for a long time and in the past has been observed also for surfaces coated with different materials, e.g.\ gold \citep{Kitagawa1976}. In fact, the mechanism seems to be mostly independent of the type of coating and rather due to the fact that electron capture by impinging protons may take place at the surface, the lower the proton energy the more efficient the process. This implies that the effectiveness of the ATHENA magnetic diverters will likely be lower than expected, which obviously translates into higher residual fluxes at the detectors. If we refer to the expected fluxes onto the WFI reported in \citealt{Fioretti2018}, the neutralisation of half of the soft protons entering the SPO at grazing incidence may be responsible for a quiet-time background level way higher than the requirement despite the magnetic shielding. Those fluxes were obtained through Geant4 \citep{Agostinelli2003} simulations based on some assumed modelisation of the diverse possible soft proton components in the space environment at L2, which however are not well known and are subject to considerable uncertainty.

In this work we follow a different approach for the assessment of
the soft proton fluxes at the ATHENA detectors, trying to derive them from the
background data measured by the German X-ray telescope \emph{extended ROentgen Survey Imaging Telescope Array} (eROSITA) (see \citealt{Predehl2021} for a comprehensive description of eROSITA and its targeted science), on board the Russian  \emph{Spektrum-Roentgen-Gamma} (SRG) spacecraft. The SRG observatory was launched in July 2019 and took for the first time X-ray instrumentation to a halo orbit around L2, therefore eROSITA looks like a very interesting and appropriate case study for this scope. Its experience in space demonstrates that for eROSITA soft protons are not a major issue, despite having no magnetic diverter against protons. During the four all-sky surveys eRASS1-4 from the end of 2019 to end of 2021 only occasionally the background level temporarily increased due to focused soft protons and in those cases it was a flaring background clearly associated with SPEs \citep{Freyberg2020}, as shown in Fig.~\ref{fig:Fig1}. 
Since the solar cycle was at its minimum in 2020, SPEs were quite rare in eRASS1. Moreover, during quiet-time no evident contribution to the background from focused soft protons has been observed, regardless of the position of the satellite along its orbit. This makes it quite tricky to quantify a potential soft proton count-rate in the eROSITA background just by comparing measurements with the filter-wheel (FW) in filter position (FWF) and in closed position (FWC), also because the two configurations differ by the FW plate in the field-of-view (FOV). 
However, we relied on a combination of measured and simulated data to obtain the result.

\begin{figure}
     \includegraphics[width=\linewidth]{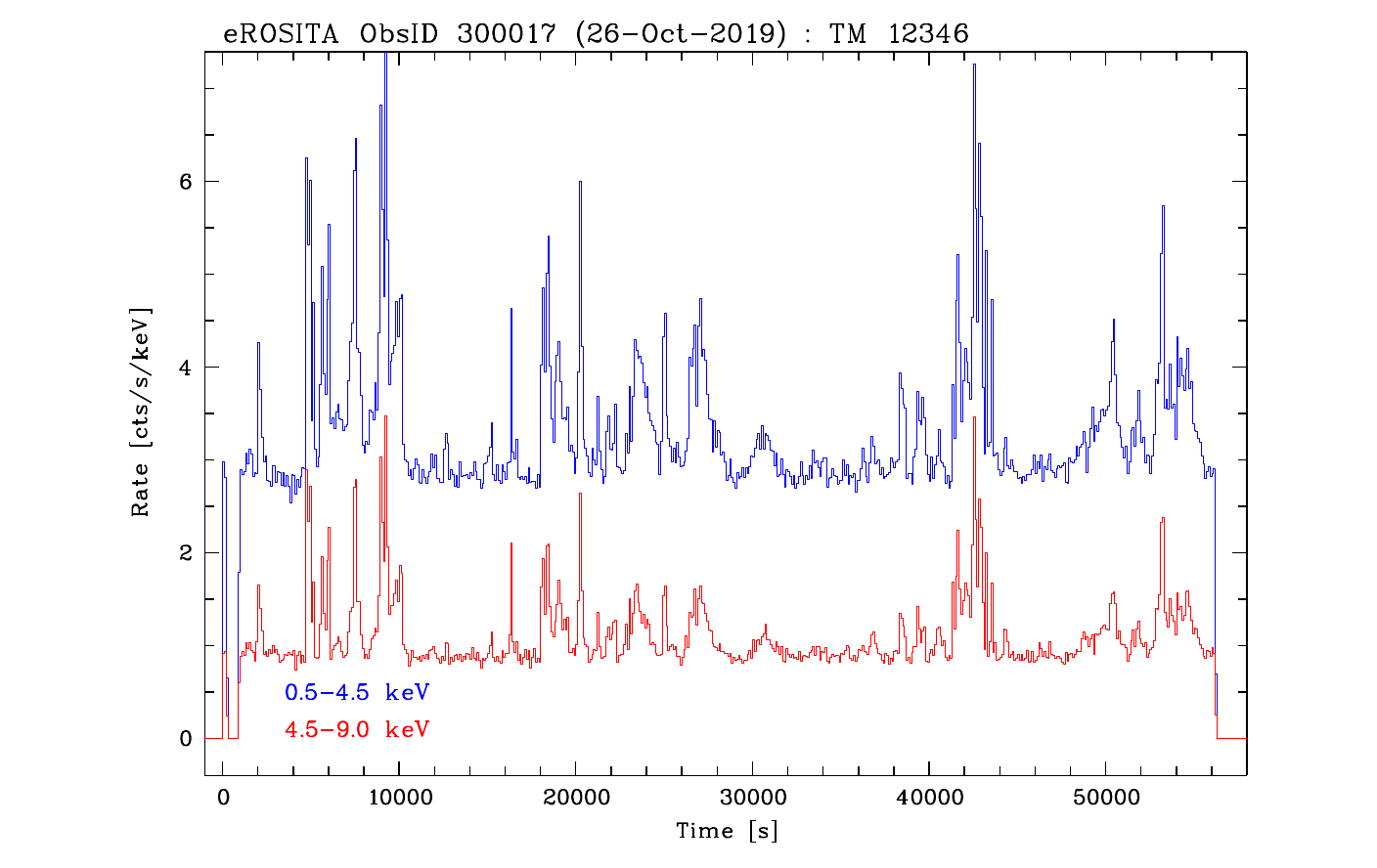}
  \caption{The first case of flaring background attributed to focused soft protons was recorded during the eROSITA PV phase, following a minor geomagnetic storm in October 2019. Soft protons affected the observation of TGU\,H2213P1 (Corona Australis dark cloud). The flaring background is shown in the 0.5-4.5 keV (\emph{blue}) and 4.5-9 keV (\emph{red}) energy bands \citep{Freyberg2020}.}
  \label{fig:Fig1}
\end{figure}

%

\section{Proton transmission yield of the eROSITA,
  WFI and X-IFU optical/thermal filters}


Each of the seven identical eROSITA Telescope Modules (TM) is equipped with
its own FW, as shown in Fig.~\ref{fig:Fig2}. An optical blocking filter (OBF) is needed for the suppression of visible/UV light. TM1, TM2, TM3, TM4 and TM6 are provided with 200 nm aluminum deposited onto the back-illuminated surface of the pnCCD at the focal plane together with a free-standing layer of 200 nm polyimide mounted on the FW. Instead TM5 and TM7 are naked (i.e.\ without on-chip aluminum layer) and just use a free-standing OBF consisting of 200 nm polyimide coated with 100 nm aluminum mounted on the FW. In FWC-position the FOV is closed by the FW plate of 4\,mm thickness, also blocking soft protons. The OBF degrades the energy of the soft protons focused onto the detector through the mirror shells. Geant4 simulations show that, for the TMs with on-chip layer, only protons reaching the OBF with energies roughly in the range 40-80 keV contribute significantly to generating background counts in the 2-7 keV band. Also for TM5 and TM7 the range is found quite similar, just shifted to about 30-70 keV. If we neglect possible small energy losses at the mirror level, the range 30-80 keV can then be expected as the most critical energy range of environmental soft protons for background generation in eROSITA.

We found a similar energy range also for the OBF currently baselined for the WFI, consisting of 90 nm aluminum deposited on-chip and a separate layer of 150 nm polyimide coated with 30 nm aluminum mounted on the FW \citep{Barbera2018a}. Assuming a flat proton spectrum in the 10-100 keV range, we show in Fig.~\ref{fig:Fig3} the comparison of the simulated proton transmission yield for the OBF of the TMs with on-chip layer and the WFI OBF (hereafter $\zeta_{TM12346}(E)$ and $\zeta_{WFI}(E)$, respectively) resulting in background generation in the 2-7 keV band.

As angular scattering inside the OBF slightly changes the directions of impinging protons, to some extent the soft proton flux onto the detector may also depend on its size as well as on its distance from the FW. This matters, in particular, for the X-IFU, that is provided with an OBF on the FW (with three or four different options depending on observational needs) paired with several fixed thermal filters mounted at different heights in a distance of 20 cm between the detector and the FW \citep{Barbera2018c}. This multi-layer configuration together with the small size of the detector favors the spread of the focused soft protons out of it. Assuming an X-ray absorber made of 1 $\mu$m gold on top of 4 $\mu$m bismuth and a detector size of $1.5 \times 1.5$\,cm${^{2}}$, we simulated the proton transmission yield of a configuration with an OBF on the FW placed at 62 cm above the detector and five identical thermal layers each made of 45 nm polyimide coated with 30 nm aluminum mounted as described in \citealt{Barbera2018c}, and we found $\zeta_{X-IFU}(E)$ < 1\% uniformly for all three baselined OBF options reported in \citealt{Barbera2018b}. 

  \begin{figure}
  \centering
    \includegraphics[width=70mm,scale=0.75]{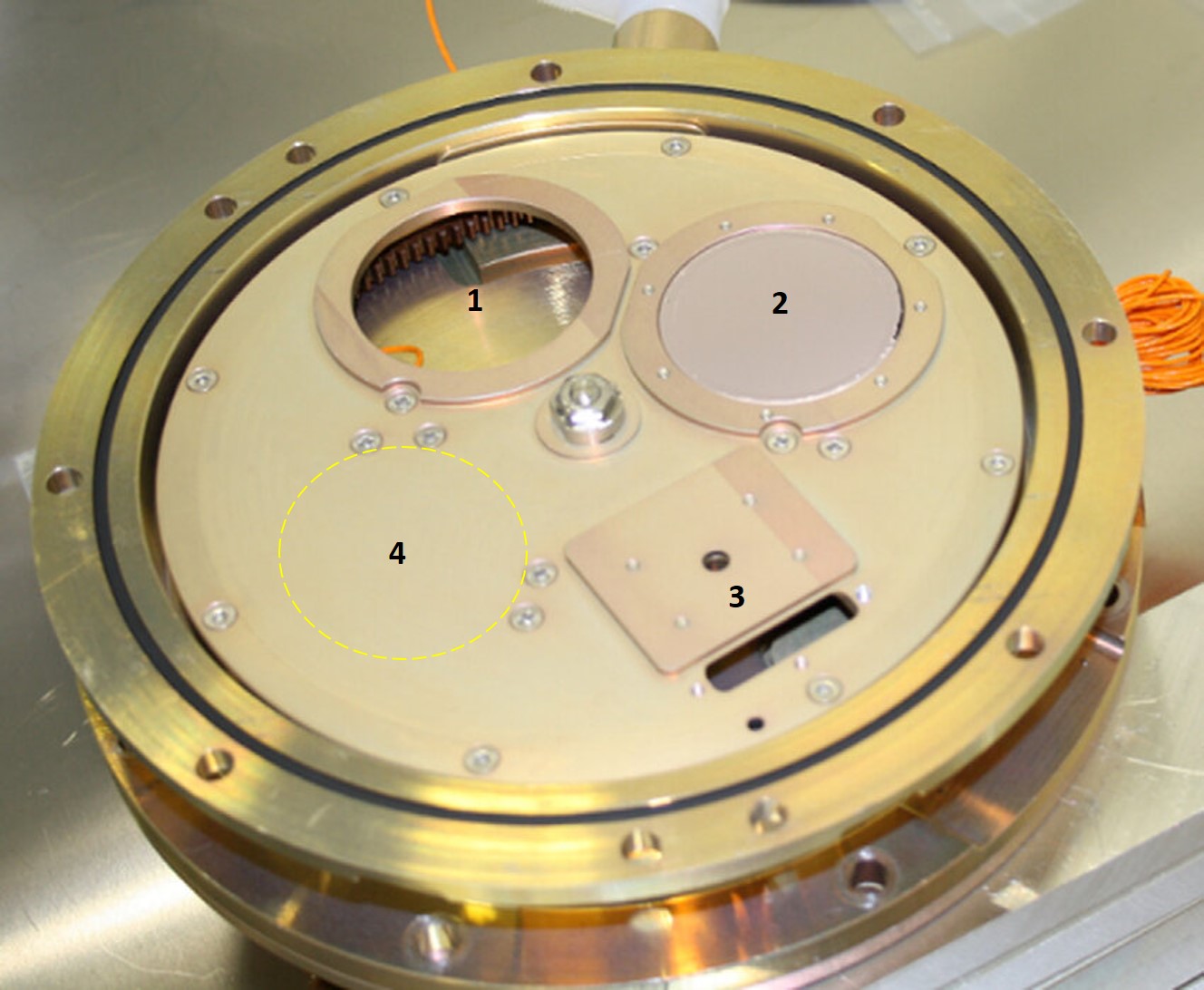}
  \caption{The four positions of the eROSITA filter-wheel are displayed: 
1:open, 2:filter, 3:calibration, 4:closed.}
  \label{fig:Fig2}
\end{figure}

\begin{figure}
  \includegraphics[width=\linewidth]{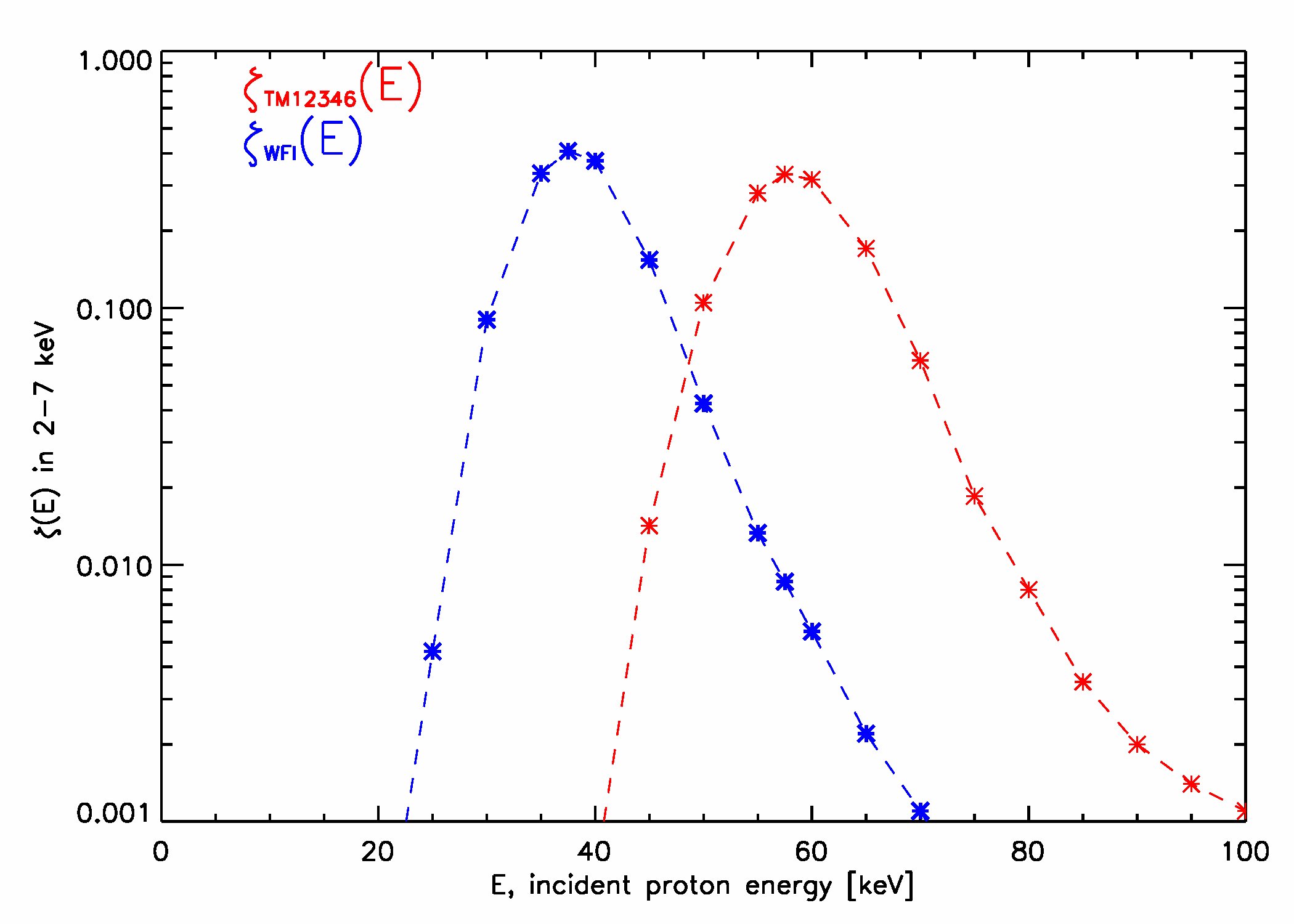}
  \caption{Simulated transmission yield $\zeta(E)$ in the 2-7 keV band of protons with incident energy in the 10-100 keV range for the eROSITA TM12346 OBF (\emph{red}) and the baseline WFI OBF (\emph{blue}). A thickness of 450\,$\mu$m is assumed for both the eROSITA pnCCD and WFI DEPFET. 
  Surface passivations of 30 nm silicon dioxide + 40 nm silicon nitride on the pnCCD and of 20 nm silicon dioxide + 30 nm silicon nitride on the DEPFET are assumed.}
  \label{fig:Fig3}
\end{figure}

   \section{Estimated soft proton count-rate in the eROSITA background. 
   Data analysis}

We investigated a possible contribution from focused soft protons in the eROSITA background data. We disregarded the few cases of enhanced background during SPEs and considered only the quiet-time background data. We also disregarded for this analysis TM5 and TM7, as they suffer from some optical leak, whose possible effects on their measured backgrounds have still to be better understood, and just exploited data from TM1, TM2, TM3, TM4 and TM6. We searched each TM individually, and in order to reduce or eliminate the systematic uncertainty related to possible small variations of the orbital fluxes on short time-scales, e.g.\ correlated with the 27-day solar rotation, for each TM we stacked data from different measurements taken during the initial performance verification (PV) phase as well as the first survey eRASS1. That means, for each TM both the used FWF-position and FWC-position datasets span a period of several months from the end of 2019 to mid 2020. We restricted the comparison to energies in the 5-7 keV band to get rid of most of the focused cosmic X-ray background (CXB) due to decreasing telescope effective area. Indeed, in FWF-position the background in this energy band is expected to be induced mainly by omni-directional energetic cosmic particles with a likely minor contribution from soft protons focused by the optics. In FWC-position only omni-directional energetic cosmic particles can contribute.  

The difference between the background in FWF-position and in FWC-position is found on the same level for all TMs, just slightly greater for TM1 and TM6 than for the others. However, hereafter we will refer to two merged datasets from the five TMs, one for all cumulated data in FWF-position and the other one for all cumulated data in FWC-position, i.e.\ we then used averaged backgrounds between the five TMs. We denote as TM8 the merging of TM1, TM2, TM3, TM4 and TM6. 

Fig. \ref {fig:Fig4} shows a comparison of the quiet-time background level measured in FWF-position and in FWC-position by TM8. The mean background level in the 5-7 keV band in FWF-position is $\sim$1\% higher than in FWC-position, i.e.\ there is an excess count-rate of $\sim$2.6·10${^{-4}}$ cts/cm${^{2}}$/s/keV. While it cannot be ruled out a priori that such a  small count-rate, or maybe just a fraction of it, may still be due to some residual CXB, for a conservative estimate we assume here that there is no contribution from the focused CXB above 5 keV. 

Furthermore, to quantify the potential soft proton count-rate we considered that in FWC-position the FW plate hit by omni-directional energetic cosmic particles likely contributes more background than the thin polyimide layer in FWF-position. Therefore, if we neglect the latter as well as any other possible unfocused background counts from within the FOV (e.g.\ from the carbon fibre panel connecting the mirror with the camera) we can conservatively add the FW plate into the calculation. Unfortunately, there is no way to disentangle its individual contribution from the overall measured background, however we can estimate it by means of Geant4 simulations. The simulated mean background from the FW plate is of $\sim$2.4·10${^{-4}}$ cts/cm${^{2}}$/s/keV, due to all omni-directional energetic cosmic particles (protons, electrons and the most abundant ions) foreseen in the interplanetary space (then we assume at L2 as well). For cosmic protons and He ions we used as an input to the simulations the \emph{solar minimum} spectra measured by PAMELA at the beginning of the last solar cycle \citep{Martucci2018}. The simulated background also amounts to $\sim$1\% of the overall measured background in FWC-position.  Therefore, we argue that in FWF-position there is a mean excess count-rate of $\sim$5·10${^{-4}}$ cts/cm${^{2}}$/s/keV. 

Error bars allow us to define lower and upper limits of this count-rate. For the lower limit, we took the lower value of the measured background in FWF-position, the higher value of the measured background in FWC-position and the lower value of the simulated background of the FW plate. Conversely, for the upper limit we took the higher value of the measured background in FWF-position, the lower value of the measured background in FWC-position and the higher value of the simulated background of the FW plate. For the measured data we only considered the statistical error (which is less than 0.5\% in both datasets) assuming that any possible systematic errors are averaged out in the same way in both datasets. For the simulated background of the FW plate we assumed an average systematic error of $\sim$10\% from the expected uncertainties on the input fluxes plus $\sim$5\% statistical error, 
i.e.\ we assigned $\sim$15\% uncertainty. 

On this basis, we calculated a minimum difference between the background level in FWF-position and in FWC-position of $\sim$3·10${^{-4}}$ cts/cm${^{2}}$/s/keV, which may be considered as a lower limit of the focused soft proton count-rate in the 5-7 keV band in TM8; and a maximum difference of $\sim$7·10${^{-4}}$ cts/cm${^{2}}$/s/keV, which may be considered as an upper limit. We expect these limits to also hold in the 2-5 keV band, where the higher CXB count-rate would complicate the quantification of any possible soft proton count-rate. 
     
\begin{figure}
  \includegraphics[width=90mm,scale=0.5]{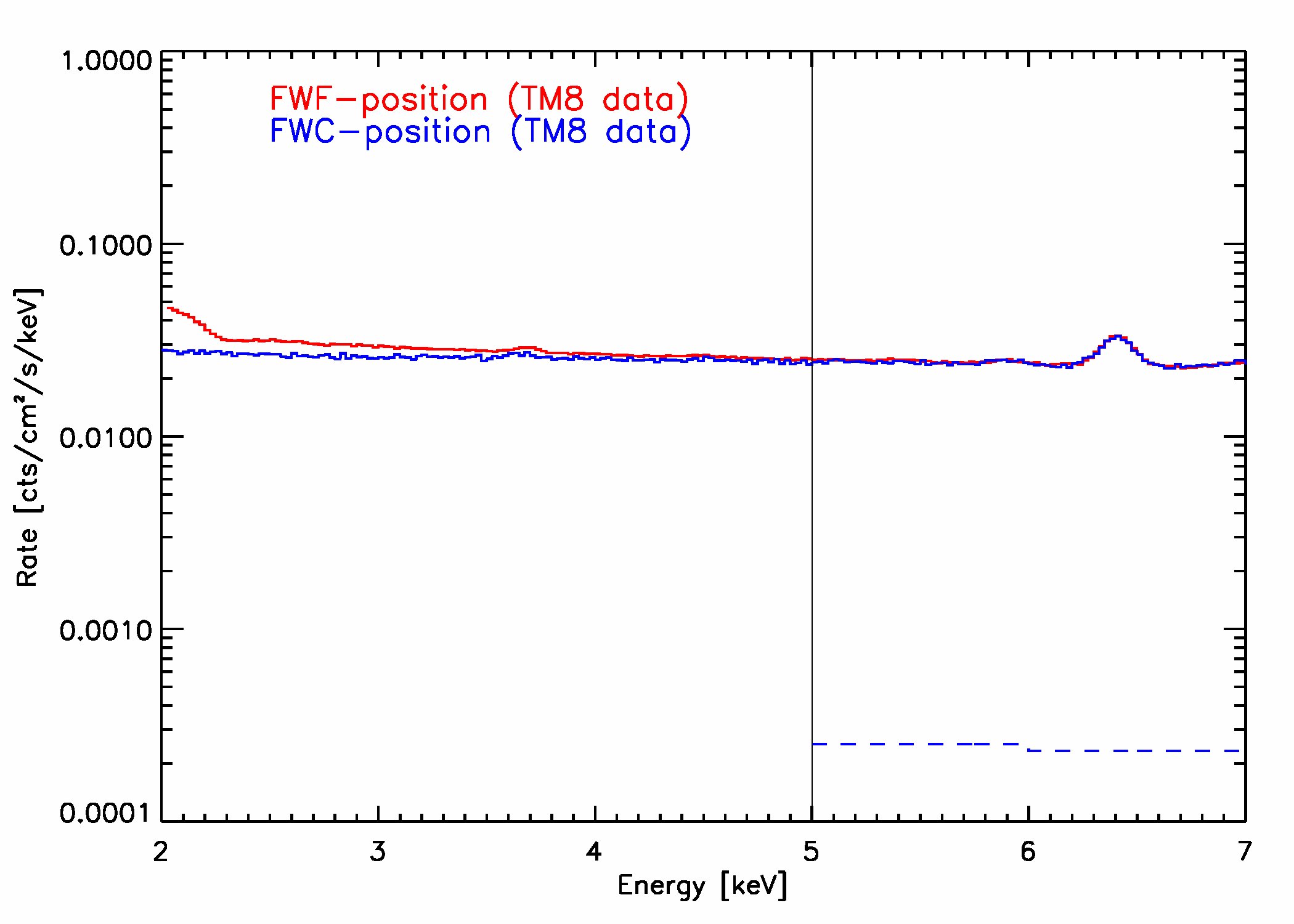}
  \caption{Comparison between the FWF-position (\emph{red}) and FWC-position
    (\emph{blue solid}) quiet-time background measured in the 2-7 keV band by the eROSITA TM8 during PV/eRASS1 (the small feature at 6.4 keV originates from iron impurities in the inner shielding). The focused CXB count-rate is not negligible below 5 keV, quite the opposite it is likely dominant over the soft proton count-rate. The simulated background from the FW plate is shown as well (\emph{blue dashed}).}
  \label{fig:Fig4}
\end{figure}

   \section{Estimated soft proton background for the WFI and X-IFU detectors}

We do not have in-situ measurements of environmental soft proton fluxes in the
SRG orbit, so we just assume ATHENA in the same orbit and try to scale the soft proton count-rate obtained for TM8 to the ATHENA detectors.
A nearly uniform proton transmission $\pi$-yield (i.e.\ the yield integrated over $\pi$) $\eta_{SPO}$ $\sim$5·10$^{-5}$ in the 10-100 keV energy range is reported in \citealt{Fioretti2018} for the SPO within a circular area with 15 cm radius at the focal point. In a previous ray-tracing study we had likewise computed for eROSITA a $\pi$-yield  $\eta_{eRO}$ $\sim$7·10$^{-4}$ \citep{Perinati2016}. However, we expect that this value is rather overestimated, as it was obtained using the Remizovich distribution \citep{Remizovich1980} to model the propagation through the mirror shells down to the focal plane of soft protons with incidence angles up to 10$^{\circ}$.

Therefore, for a consistent comparison with the SPO $\pi$-yield reported in \citealt{Fioretti2018}  we implemented a model of the eROSITA telescope in Geant4. We scaled the 15 cm radius at the SPO focal point by a factor 7.5, corresponding to the focal length (FL) ratio (FL$_{SPO}$=12 m, FL$_{eRO}$=1.6 m), and computed using the \emph{single scattering} physics list the $\pi$-yield within a circular area with 2 cm radius at the eROSITA focal point, and we obtained $\eta_{eRO}$ $\sim$10${^{-4}}$. Assuming to the first order a similar response to soft protons of the eROSITA and SPO mirror coatings (gold and iridium, respectively), as our laboratory tests in fact have shown \citep{Amato2021}, by geometry the $\pi$-yield of the eROSITA optics could be expected $\sim$25\% higher than that of the SPO, its aperture being $\sim$45 times smaller while the scaled circular area at its focal point being $\sim$56 times smaller (for the assumed apertures see Table 1 at the end of this section). As the $\pi$-yield increases for larger effective solid angles, the simulated enhanced value for eROSITA can be explained with a proton acceptance angle somewhat larger than that of $\sim$5$^{\circ}$ reported in \citealt{Fioretti2018} for the SPO, the eROSITA FOV (61 arcmin diameter) being larger than the SPO FOV (40 arcmin diameter).

We derive the soft proton induced WFI and X-IFU backgrounds (SPB) from the estimated soft proton induced TM8 background with the following scaling-laws:      
\begin{equation}
       SPB_{WFI} \sim SPB_{TM8} \times \frac{\text{$\eta_{SPO}$}}{\text{$\eta_{eRO}$}} \times \frac{\text{$\zeta_{WFI}$}}{\text{$\zeta_{TM8}$}} \times \frac{\text{$\Omega_{WFI}$}}{\text{$\Omega_{SPO}$}} \;,
   \label{Eq-1}
\end{equation}
\begin{equation}
   SPB_{X-IFU} \sim SPB_{WFI} \times \frac{\text{$\zeta_{X-IFU}$}}{\text{$\zeta_{WFI}$}} \times \frac{\text{$\Omega_{X-IFU}$}}{\text{$\Omega_{WFI}$}}  
   \;,
   \label{Eq-2}
\end{equation}
 
\noindent
where $\eta_{eRO}$ and $\eta_{SPO}$ are the proton transmission $\pi$-yields of the eROSITA and SPO optics, respectively; $\zeta_{TM8}$ (=$ \zeta_{TM12346}$), $ \zeta_{WFI}$ and $ \zeta_{X-IFU}$ are the probabilities (i.e.\ the yields $ \zeta(E)$ integrated in the 10-100 keV energy range) of transmitting background in the 2-7 keV band of the eROSITA TM8, ATHENA WFI and X-IFU OBFs, respectively; 
$\Omega_{SPO}$, $ \Omega_{WFI}$ and $ \Omega_{X-IFU}$ are the proton acceptance solid angles of the SPO, WFI and X-IFU, respectively.

\begin{table}
\caption{Assumed mirror and detector apertures for the scaling in Eqs.(1) and (2), for ATHENA and eROSITA TM8 (single telescope).} 
\label{table:1} 
\centering 
\begin{tabular}{c c } 
\hline\hline 
\tiny Aperture & \tiny Geometric area [cm${^{2}}$]  \\ 
\hline 
$A_{SPO}$   & 43115 \\
$A_{eRO}$   & 954  \\
$A_{WFI}$   & 225  \\
$A_{X-IFU}$ & 2.25 \\
$A_{pnCCD}$ & 8.3  \\
 \hline
\end{tabular}
\end{table}

 \subsection{Estimated soft proton background for the WFI}
For the WFI, which covers the whole FOV (40 arcmin diameter), we assume the same
proton acceptance angle as for the SPO, i.e.\ $\Omega_{WFI}$ = $\Omega_{SPO}$. 
Moreover, it is $\zeta_{WFI}$ $\sim$ $\zeta_{TM8}$, as shown in Fig.~\ref{fig:Fig3}, then in principle the dependence on $\zeta$ in Eq.(1) would cancel out as well. However, $\zeta(E)$ in 
Fig.~\ref{fig:Fig3} has been computed for a flat proton spectrum, and the fact that $\zeta_{WFI}(E)$ peaks at somewhat lower proton energies than $\zeta_{TM8}(E)$ may point to a less favorable configuration, as the spectra of the known soft proton sources in space are expected to follow power-laws decaying with energy. As an example, we simulated the passage of the solar wind input spectrum reported in \citealt{Fioretti2018} through both the TM8 and WFI OBFs, in that case $\zeta_{WFI}$ $\sim$2·$\zeta_{TM8}$. That means, if that spectrum corresponded to the real one encountered by SRG (which we do not know), given that $\eta_{SPO}$ $\sim$1/2·$\eta_{eRO}$ from Eq.(1) we could actually expect a soft proton induced WFI background approximately in the same range we estimated in section 3 for TM8, i.e.\ 3·10${^{-4}}$< $SPB_{WFI}$< 7·10${^{-4}}$ cts/cm${^{2}}$/s/keV, where the upper limit lies just slightly above the requirement. 

Intuitively, to really drop the dependence of the scaling Eq.(1) on $\zeta$ a change to the baseline WFI OBF might be helpful, namely to make its specifications the same as those of the TM8 OBF. In this case, in fact, whatever the external soft proton fluxes in the SRG orbit, the scaling Eq.(1) always gives $SPB_{WFI}$ $\sim$1/2·$SPB_{TM8}$, i.e.\ 1.5·10${^{-4}}$< $SPB_{WFI}$< 3.5·10${^{-4}}$ cts/cm${^{2}}$/s/keV, well within the requirement. 

Although this change would work to lower the soft proton induced background in the 2-7 keV band, it would have as a drawback an impact on the quantum efficiency at low energies, so probably it would not be worth it, if the soft proton induced WFI background was actually in the low count-rate regime estimated here. Nevertheless, it could still be taken into consideration as a precaution. As a compromise, just one of the two layers could possibly be made somewhat thicker, and in this case increasing the amount of aluminum would be more effective for reducing the soft proton flux than thickening the layer of polyimide. For instance, we compare in Fig.~\ref{fig:Fig5} the X-ray transmissivity of the baseline WFI OBF and that of a similar OBF where the thickness of the aluminum coating has been doubled from 30 nm to 60 nm. On the one hand, the detector response below 0.5 keV would be to some extent sacrificed; on the other hand, we point out that, although no specific requirement was set for the low energy background induced by soft protons, should the planned magnetic diverters be only partially effective the 0.2-2 keV band could also be affected by a higher soft proton flux, 
e.g.\ an enhanced background at the lower energies due to focused soft protons is reported in \citealt{Kuntz2008} for XMM-Newton (though in that case a flare spectrum is described). Thence, a somewhat thicker WFI OBF might still deliver a better signal-to-noise ratio even below 0.5 keV despite its lower X-ray transmissivity, the optimisation of this trade-off clearly depending on the actual soft proton flux expected in that band. 

\begin{figure}
\centering
  \includegraphics[width=\linewidth]{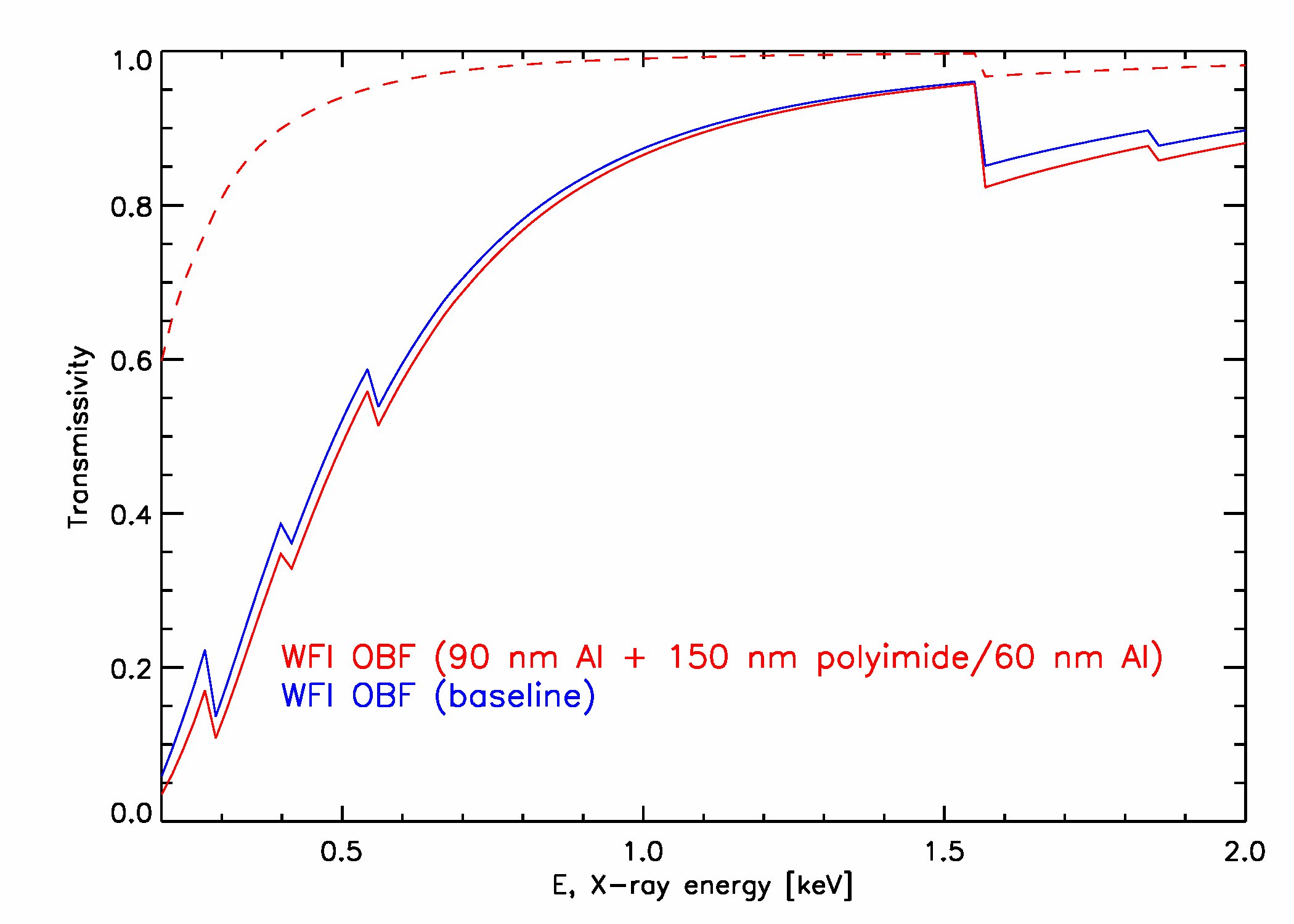}
  \caption{Computed X-ray transmissivity of a hypothetical WFI OBF with thicker (60 nm) aluminum coating (\emph{red solid}) and the baseline WFI OBF (\emph{blue}) in the 0.2-2 keV band. The transmissivity ratio (\emph{red dashed}) highlights the lower transmissivity of the OBF with the thicker coating.}
  \label{fig:Fig5}
  \end{figure}

\subsection{Estimated soft proton background for the X-IFU}

For the X-IFU, which covers a smaller FOV (5 arcmin diameter), we assume a
proton acceptance angle of $\sim$1.5$^{\circ}$, i.e.\ a solid angle ratio
$\Omega_{X-IFU}$/$\Omega_{WFI}$ $\sim$0.1.
This angular ratio and the fact that $\zeta_{X-IFU}$<$\zeta_{WFI}$, 
as discussed in section 2, make the X-IFU configuration less susceptible to soft protons. Therefore, for the X-IFU the scaling Eq.(2) nicely gives for all baselined OBF options a soft proton induced background within the requirement.\\ 



\section{Final remarks}

It is worth stressing that the mentioned values of the parameters assumed for the scaling as well as the soft proton background requirement of 5·10${^{-4}}$ cts/cm${^{2}}$/s/keV in the 2-7 keV band refer to the former ATHENA configuration. In the new ATHENA configuration recently approved by ESA some changes are envisaged, in particular the number of the SPO rows will be reduced from 15 to 13. Since  this change goes in the direction of lowering the proton transmission yield of the SPO, the estimated fluxes in section 4 can be considered as conservative, at least as far as the design of  the OBFs does not undergo major changes. \\

\section{Summary and conclusions}
The German telescope eROSITA on board SRG provided the first measurements of an X-ray instrument in orbit around the second Lagrangian point L2 and we tried to use its background data to make a prediction of soft proton induced background for the ATHENA detectors, independently of any modelisation of the external soft proton fluxes in the space environment and just assuming ATHENA in the same orbit as SRG.  

First, we defined a range of values for the potential soft proton count-rate in the eROSITA background data. Then we scaled it to the ATHENA detectors and found that, even without magnetic field, both the soft proton induced WFI and X-IFU backgrounds (in FWF-position) could be expected close to the requirement. 

Of course, this encouraging result cannot be generalized to every orbit in space, hence we underline the importance of ATHENA being equipped with the planned magnetic diverters and strongly recommend their application. In order to make up for the possible incomplete response of the magnets from the expected neutralisation of the soft protons scattered by the SPO, a mitigation strategy based upon increasing somewhat, e.g.\ by $\sim$30\%, the thickness of the OBF baselined for the WFI, which may be affected by a significantly higher soft proton flux than the X-IFU, is suggested to shift its critical range towards slightly higher proton energies. However, since this would be achieved at the cost of some undesirable degradation of the low energy quantum efficiency, the pros and cons of this change must be investigated more thoroughly and weighed up very carefully.

\begin{acknowledgements}
This work is based on data from SRG/eROSITA, the soft X-ray instrument aboard SRG, a joint Russian-German science mission supported by the Russian Space Agency (Roskosmos), in the interests of the Russian Academy of Sciences represented by its Space Research Institute (IKI), and the Deutsches Zentrum für Luft- und Raumfahrt (DLR). The SRG spacecraft was built by Lavochkin Association (NPOL) and  its subcontractors, and is operated by NPOL with support from the Max Planck  Institute for Extraterrestrial Physics (MPE). 
The development and construction  of the SRG/eROSITA X-ray instrument was led by MPE, with contributions from the Dr. Karl Remeis Observatory Bamberg and ECAP (FAU Erlangen-Nuernberg), the University of Hamburg Observatory, the Leibniz Institute  for Astrophysics Potsdam (AIP), and the Institute for Astronomy and Astrophysics  of the University of Tübingen, with the support of DLR and the Max Planck Society. The Argelander Institute for Astronomy of the University of Bonn and the Ludwig Maximilians Universität Munich also participated in the science preparation for SRG/eROSITA.\\
The authors acknowledge support by the DLR under grant no. 50 QR 2102 and 50 QR 2302.
\end{acknowledgements}

%
%
\bibliographystyle{aa} 
\bibliography{dr1} 
\end{document}